\documentstyle[12pt]{article}

\def\dspace{\baselineskip = .30in}

\begin{document}

\def\ie{\hbox{\it i.e.}{}}      \def\etc{\hbox{\it etc.}{}}
\def\eg{\hbox{\it e.g.}{}}      \def\etal{\hbox{\it et al.}}

\def\simgt{\ \raisebox{-.25ex}{$\stackrel{>}{\scriptstyle \sim}$}\ }
\def\simlt{\ \raisebox{-.25ex}{$\stackrel{<}{\scriptstyle \sim}$}\ }
\def\grad#1{\bigtriangledown #1}
\def\VEV#1{\left\langle #1\right\rangle}

\def\UN#1{\ {\rm #1}}
\def\TeV{\UN{TeV}}
\def\GeV{\UN{GeV}}
\def\MeV{\UN{MeV}}
\def\eV{\UN{eV}}
\def\keV{\UN{keV}}
\def\abs#1{\left| #1\right|}

\def\appeqs{\setcounter{equation}{0}
\def\theequation{A\arabic{equation}}}

\def\omz{\omega_0}
%%%%%%%%%%%%%%%%%%%%%%%%%%%%%%%%%%%%%%%%%%%%%%%%%%%%%%%%%%%%%%%%%%%%%%%%

\begin{titlepage}
\begin{flushright}
BA-94-21  \\
\today
\end{flushright}
%\par
\vspace{.3in}
\begin{center}
{\Large{\bf Axion Dissipation Through the Mixing of Goldstone Bosons}}\\
\vskip 0.3in
{\bf K.S. Babu, S.M. Barr and D. Seckel}\\[.2in]
{\it Bartol Research Institute \\
University of Delaware, Newark, DE 19716\\}

\end{center}
\vskip 0.4in

\begin{abstract}
By coupling axions strongly to a hidden sector, the energy density in
coherent axions may be converted to radiative degrees of freedom,
alleviating the ``axion energy crisis''. The strong coupling is achieved
by mixing the axion and some other Goldstone boson through their
kinetic energy terms, in a manner reminiscent of paraphoton models.
Even with the strong coupling it proves difficult to relax
the axion energy density through particle absorption, due to the
derivative nature of Goldstone boson couplings and the effect of back
reactions on the evolution of the axion number density. However, the
distribution of other particle species in the hidden sector will
be driven from equilibrium by the axion field oscillations. Restoration
of thermal equilibrium results in energy being transferred
from the axions to massless particles, where it can
redshift harmlessly without causing any cosmological problems.
\end{abstract}

%\begin{flushleft}
%Pacs numbers:  14.80.Pb, 97.60.Bw, 25.90.+k
%\end{flushleft}
\end{titlepage}

%%%%%%%%%%%%%%%%%%%%%%%%%%%%%%%%%%%%%%%%%%%%%%%%%%%%%%%%%%%%%%%%%%%%%%%%

\dspace

\section{Introduction}
Goldstone and pseudo--goldstone bosons play a large role in the
speculations of particle physicists and cosmologists.  Some
oft--discussed examples are axions,$^1$ majorons,$^2$ familons,$^3$ and
the non--abelian goldstone fields of texture models.$^4$  The salient
feature of the dynamics of goldstone bosons is that they couple
derivatively.  If the associated global continuous symmetry is broken at
a scale $f$, then the coupling always involves powers of
$f^{-1}\partial_\mu b$, where $b$ is the goldstone boson.  The consequence
of this is that if $f$ is very large compared to the relevant masses and
momenta, the goldstone boson decouples.  This fact is exploited in
invisible axion models,$^5$ where for $f_a \stackrel{_>}{_\sim} 10^{10}\GeV$
the axion's
couplings are small enough for it to have evaded detection and avoided
conflict with astrophysical observations.$^6$  This fact also lies
behind the well--known ``axion energy problem'',$^7$ which is that if
$f_a \stackrel{_>}{_\sim} 10^{12}\GeV$ the energy in
coherent oscillations of the axion
field after the QCD phase transition in the early universe is unable to
dissipate due to the axion's weak coupling and eventually overcloses the
universe.

In this letter we show that different goldstone fields can mix with each
other through kinetic terms in the Lagrangian in a way reminiscent of
photon--paraphoton mixing,$^8$ and that a goldstone boson can thus
acquire effective couplings at low energy to certain fields
which are much larger than the
typical $p/f$.  After demonstrating this point we show how it makes
possible scenarios in which the energy in coherent oscillations of the
cosmological axion field can be dissipated and the upper limit on $f_a$
removed.

\section{Mixing of goldstone bosons}

Consider a theory with a global symmetry $U(1)_a \times U(1)_b$.  Let
$\Omega$ and $\Phi$ be two scalar fields with charges (1,0) and (0,1)
under this symmetry.  Suppose that $\abs{\VEV{\Omega}} \equiv f_a$ and
$\abs{\VEV{\Phi}} \equiv f_b$, where $f_b$ is assumed to be much smaller than
$f_a$, and call the resulting goldstone modes of $\Omega$ and
$\Phi$, `$a$' and `$b$' respectively.  Then it is clear that couplings of
$a$ will be suppressed by $p/f_a$ and those of $b$ by $p/f_b$.

It can happen that by integrating out some fields of mass $M$ an
effective higher dimension operator will result of the
$U(1)_a \times U(1)_b$--invariant form
\begin{equation}
{\cal L}_{\rm mix} = {c \over {M^2}} \Omega^{\dagger} \partial_\mu
\Omega \Phi^{\dagger}\partial^\mu \Phi + h.c.
\end{equation}
If  $\Omega = (f_a+\tilde{\Omega})e^{ia/f_a}$ and
$\Phi = (f_b+\tilde{\Phi})e^{ib/f_b}$ are substituted into this
expression, the result contains the term
\begin{equation}
{\cal L}_{a,b} = \epsilon \left(\partial_\mu a \partial^\mu b\right)
\end{equation}
with
\begin{equation}
\epsilon \equiv -2c{{f_af_b}\over {M^2}}~.
\end{equation}
Through this mixing of $a$ and $b$, the `axion' $a$ can couple to
particles which have $U(1)_b$ charges with effective strength
$\epsilon p/f_b$, which can be much greater than $p/f_a$
(see Fig. 1).

For example, a term such as in Eq. (1) could arise from integrating out
a field $\eta$, with charges $(-1/2, -1/2)$, mass $M$, and coupling
$[\lambda (\Omega \Phi \eta^2)+h.c.]$.  For simplicity let $\eta$ have
positive mass-squared and no VEV.  Then the loop shown in Fig. 2 will
produce the following effective term (among others) for momenta less
than $M$:
\begin{equation}
{{\lambda^2}\over{192 \pi^2 M^2}}
(\Omega^{\dagger}\partial_\mu \Omega \Phi^{\dagger}
\partial^\mu \Phi + h.c.)
\end{equation}
and therefore, from Eq. (3),
\begin{equation}
\epsilon = -{{\lambda^2}\over {96 \pi^2}} {{f_af_b}\over {M^2}}~.
\end{equation}
Now, given that the term $\lambda(\Omega \Phi \eta^2)$ exists, it would
require an artificial cancellation for $M^2$ to be much less than
$\lambda f_af_b$, and hence for $\epsilon$ to be much larger than
$\lambda/(96\pi^2) \sim 10^{-3}$.  There is also an absolute limit
that $\epsilon < 1$, since the kinetic terms of $a$ and $b$ have the
form
$$
{1 \over 2} \left(\matrix{\partial_\mu a & \partial_\mu b}\right)
\left(\matrix{1 & \epsilon \cr \epsilon & 1}\right)\left(\matrix{
\partial^\mu a \cr \partial^\mu b}\right)
$$
and a wrong--sign metric would otherwise result.  There is no reason,
however, why $\epsilon$ cannot be of order, though less than, unity.

If the higher dimension term in Eq. (1) were a ``hard'' term all the way
up to momenta of order $f_a$,  then the theory would become strongly
coupled and physics uncalculable unless $M^2$ were $\ge f_a^2$ and
hence $\epsilon$ were less than about $f_b/f_a$.  The effective coupling
of $a$ to particles with $U(1)_b$ charges through mixing then would be
of order $\epsilon p/f_b \le p/f_a$.  In other words, the coupling
of $a$ would not
be enhanced by this mixing.  If, on the other hand, the term in Eq. (1)
arises through a loop such as that shown in Fig. 2 where the
intermediate particles have mass $M^2 \sim f_af_b$, then it softens for
momenta above $M$ and no problem arises in higher order.

\section{Alleviating the axion energy crisis}

We wish to make use of the mixing of goldstone
bosons to let axions couple strongly to some other particles and
thus, somehow, allow the primordial energy density in axions to be dissipated.

Let the field $a$ be the
axion that solves the strong CP problem.  Then $U(1)_a$ has a QCD
anomaly and $a$ gets a mass of order $\Lambda_{QCD}^2/f_a$ from
instanton effects. Several astrophysical arguments$^6$ suggest that
$f_a \simgt 10^{10}\GeV$, otherwise the evolution of stars or their
remnants would be drastically different than observed.
Thus the coupling of the axion to ordinary matter,
which is proportional to $1/f_a$, is highly suppressed.

The axion energy crisis arises as a result of the small coupling and the
small mass of the axion. At temperatures above the QCD phase transition
the axion mass is small enough that $\dot{a} \approx 0$.
As the universe cools, $m_a$ increases until the axion field becomes
dynamic, \ie\ $m_a > H$, where $H$ is the expansion rate. Barring a
coincidence of alignment for the Peccei-Quinn angle, the initial
value of the axion field is $a \sim f_a$. The axion number density
is then $n_a \approx f_a^2 H$. After that point the number
of axions in a comoving volume is constant, while the axion mass increases
from $H$ to its zero temperature value $m_{a0}$. If the energy density
of axions after the QCD phase transition exceeds a critical value,
$\rho_{a,cr} \approx 10^{-7} \Lambda_{QCD}^4$, then the
the universe would become matter dominated too early, \ie\ at temperatures
$\simgt 10\eV$. The energy density in axions at $T\sim \Lambda_{QCD}$
is roughly $\rho_a(\Lambda_{QCD}) \sim \Lambda_{QCD}^4 f_a/M_{Pl}$, which
results in the constraint $f_a < 10^{12} \GeV$.
In non-inflationary
models there are bound to be axionic strings - whose decay may$^{9}$
or may not$^{10}$ increase
the axion abundance 100-fold, lowering the cosmological bound to
$f_a \simlt 10^{10}\GeV$.

There have been several approaches to solving
the axion energy crisis. For example, entropy production$^{11}$
may dilute the energy density in axions relative to ordinary
matter, or inflation may fix the initial value of the
Peccei-Quinn angle to a suitably small value$^{12}$. However,
attempts to dissipate the axion energy
typically fail. Direct axion absorption presents several
challenges. First, the goldstone boson nature of axions demands a factor
of $1/f_a$ at each vertex. Second, since the axions to be absorbed are
non-relativistic the derivative couplings demand an
additional factor of $m_a$ at each vertex. Third, any process
for single axion absorption must admit axion emission in the time
reversed channel. This leads to a reduction in the dissipation rate by
a factor of $m_a/T$ from a naive approach that would ignore the back
reactions. Flynn and Randall$^{13}$ considered the non-derivative coupling
of axions to mesons$^{14}$, but the first and third arguments above were
sufficient to leave axions undamped by several orders of magnitude.

In the present case, the mixing of axions with some other goldstone boson
allows the possibility of coupling axions strongly to some other
non-standard matter. However, without some explicit symmetry breaking of
the $U(1)_b$, the axions must couple derivatively. The combination of
powers of $m_a$ from the couplings and from consideration of back reactions
has caused all our attempts at axion absorption to fail.

Nonetheless, there is a mechanism whereby axion oscillations can be
damped, allowing for $f_a$ of order the grand unified scale $M_{GUT}$,
or even the Planck scale $M_{Pl}$. Let the axion be mixed,
as described above, with strength $\epsilon$ with a goldstone boson
$b$ arising from the breaking of $U(1)_b$.
The axion coupling to `$b$--matter', as we will
call it, will be vastly stronger than its coupling to ordinary matter.
Our basic idea is then that the oscillating axion field causes
the energies of $b$-particles to change in energy. If a thermal bath of
such $b$-particles were present in the early universe when the axion
field started to oscillate, then those oscillations would drive
the $b$-distribution out of thermal equilibrium. By adjusting the
interaction rates of the $b$-particles to a suitable value the axion energy
can be dissipated into thermal energy of $b$ particles. The final step is
to note that the $b$ goldstone boson is massless, so that all $b$ energy
can in principle redshift away as radiation. The axion energy is thus put
into a harmless form which never dominates the universe.

\section{An explicit model}

To illustrate how this may happen
let us calculate in a simple model in which the
`$b$--matter' consists of the field $\Phi$ whose vacuum expectation
value breaks $U(1)_b$ and whose phase is $b$, and a scalar field
$\chi$ which has no VEV.  Let the $U(1)_b$ charges of $\Phi$ and $\chi$
both be $+1$ and let them have a coupling $({1 \over 4} g \Phi^{*^2}
\chi^2 + h.c.)$.

The terms of the Lagrangian that will be relevant for our analysis are
\begin{eqnarray}
{\cal L} &=& {1 \over 2} (\partial_\mu a)^2 - {1 \over 2} m_a^2 a^2 +
{1 \over 2} (\partial_\mu b)^2 + \epsilon (\partial_\mu a \partial^\mu b) +
\nonumber \\
& ~& {1 \over 2} |\partial_\mu \chi |^2 - {1 \over 2} m^2|\chi |^2 -
{1 \over 4} \mu^2 \left( e^{-2ib/f_b}
\chi^2+h.c.\right).
\end{eqnarray}
The last term arises from expanding $\Phi$ as $(f_b+\tilde{\Phi})e^{ib/f_b}$
in the $\Phi^{*^2} \chi^2$ coupling and defining
$\mu^2\equiv gf_b^2$. At this point we are free to choose the parameters
of ${\cal L}$. For purposes of illustration it
is convenient to choose $f_b \sim m \sim \Lambda_{QCD}$ and to choose
$\mu$ somewhat less than $m$. These values are not absolutely necessary,
but they clearly show the damping mechanism to work.

The first task is to resolve the mixing of $a$ and $b$.
To bring the kinetic terms of $a$ and $b$
to canonical form we perform a non--orthogonal transformation
\begin{eqnarray}
a^\prime &=& a\sqrt{1-\epsilon^2} \nonumber \\
b^\prime &=& b+\epsilon a~.
\end{eqnarray}
This gives
\begin{eqnarray}
{\cal L} &=& {1 \over 2} ({\partial_\mu} a^\prime)^2 -{1 \over 2} m_a^2 (1-
\epsilon^2)^{-1} a^{\prime^2} + {1 \over 2} (\partial_\mu b^\prime)^2
\nonumber \\
&+& {1 \over 2} |\partial_\mu \chi |^2 -{1 \over 2} m^2 |\chi |^2
-{1 \over 4} \mu^2 \left(
e^{2i/f_b(-b^\prime+{{\epsilon a^\prime} \over {\sqrt{1-\epsilon^2}}})}
\chi^2+h.c.\right) ~.
\end{eqnarray}

At this point it is convenient to drop the primes and work with the
canonically normalized fields. For further convenience we represent the
coherent oscillation of the axion field by
\begin{equation}
a(t) = a_0 \sqrt{1-\epsilon^2} \sin (m_a t)~.
\end{equation}
Then the effective Lagrangian for $\chi$ in the presence of axion
oscillations becomes
\begin{equation}
{\cal L}_{eff} = {1 \over 2} |\partial_\mu \chi |^2 -{1 \over 2} m^2
|\chi |^2 -
{1 \over 4} \mu^2\left( e^{2i \theta(t)}\chi^2+h.c.\right)
\end{equation}
with $\theta(t) \equiv (\epsilon a_0/f_b) {\rm sin} (m_a t)$.

It is important to note that even though $a(t)$ is slowly
varying (with frequency $m_a$), $\theta(t)$ is rapidly varying:
$q \equiv \dot{\theta}(t) = \epsilon {{a_0 m_a} \over {f_b}}
{\rm cos}(m_at)$.  Since $a_0m_a = \rho_{\rm axion}^{1/2} \sim
\Lambda_{QCD}^2 (f_a/M_{pl})^{1/2}$ (initially), $q \sim \epsilon
\Lambda_{QCD}$.  For
time scales short compared to $m_a^{-1}$ we may treat $q$ as being
constant and write ${\cal L}$ as
\begin{equation}
{\cal L}_{eff} = {1 \over 2} |\partial_\mu \chi |^2 - {1 \over 2}
m^2|\chi |^2
-{1 \over 4} \mu^2 \left[e^{2iq t}\chi^2 + h.c\right]~.
\end{equation}

The problem now is to find the evolution of the $\chi$ field, a
non-trivial task because of the explicit time dependence. This is done in
the Appendix, but we give a simplified version here.
First, if $\mu^2 =0$ then for each momentum $\vec{k}$ there would be two
degenerate eigenstates. One could choose to work in a basis with
linearly polarized or with circularly polarized states. With degenerate
states it is easy to show that the Hamiltonian is constant for any linear
combination. Next, consider the case $\mu^2 \ne 0$, but
$q = 0$. Then the two eigenstates have masses
$\left(m^2 \pm \mu^2\right)^{1 /2}$, and correspond to the linearly
polarized states along the real and imaginary components of $\chi$;
\ie\ they are $A(x^\mu)$ and $B(x^\mu)$, where
$\chi(x^\mu) = A(x^\mu) + iB(x^\mu)$. Now, there is no explicit
time dependence so the Hamiltonian is still constant. This can be seen
explicitly by writing the Hamiltonian in terms of the linearly polarized
states and observing that there is no mixing of different frequency states
in the Hamiltonian. Finally, for non--zero $q$, the Hamiltonian is time
dependent and the energy of the $\chi$ field varies with time.
For $\mu^2$ small the evolution is well approximated by the $\mu^2 = 0$
eigenstates, but there is a qualitative difference between the linearly
polarized and circularly polarized solutions.

For the linearly polarized case, the potential in
the $A$ (or $B$) direction has a curvature that varies sinusoidally
in time: $m_{eff}^2 = m^2 \pm \mu^2 {\rm cos}(2q t)$. Since we will assume
$q \ll m$, the variation in  $m^2_{eff}$ is adiabatic.
In that case one expects each quantum to have energy
$\sqrt{m^2 + \vec{k}^2 +\mu^2 {\rm cos}(2q t)} \simeq
\omz + {1 \over 2} {{\mu^2}\over {\omz}}
{\rm cos}(2q t)$, where $\omz \equiv \sqrt{m^2 + \vec{k}^2}$.
The number of quanta of oscillation is an adiabatic invariant.

On the other hand, for the circularly polarized case, the state samples
the potential at all phases of $\chi$. Since $m \gg q$, this averaging
takes place in a time small compared to the evolution of the potential and
there is little time dependence in the energy. In the Appendix we
show that, in fact, the true eigenmodes are approximately circularly
polarized.
There is no time dependence for pure states; however, in a thermal bath we
do not expect pure states unless there is some conserved quantum number.
The corresponding symmetry in our case is the $b$ symmetry, which is
broken, so that $b$ charge is not conserved. We therefore expect all
polarizations to play a role, and having identified the linearly polarized
states as having an energy dependence driven by the axion field we look
for dissipation of the axion energy through those modes.

Time variation of the energy levels of the $\chi$ field can
lead to dissipation.
Suppose the rate at which a linearly polarized $\chi$ particle of
momentum $\vec{k}$ scatters is given by $\gamma(\vec{k})$.  Then the
occupation number $f_{\vec{k}}$, of that state will obey the
differential equation
\begin{equation}
\dot{f}_{\vec{k}} = \gamma(\vec{k}) \left[f_0\left({{E(\vec{k},t)} \over
T}\right) - f_{\vec{k}}(t)\right]
\end{equation}
where $E(\vec{k},t) = \omz + {1 \over 2} {{\mu^2}\over
{\omz}} {\rm cos}(2q t)$ and $f_0$ is the equilibrium
occupation number (Bose--Einstein).  Treating ${{\Delta E}\over T} =
{1 \over 2} {{\mu^2}\over {T\omz}}$ as small, this has
solution
\begin{equation}
f_{\vec{k}}(t) = f_0\left({{\omz}\over T}\right) +
f_0^\prime \left({{\omz}\over T}\right)
{\gamma \over {\sqrt{\gamma^2+4q^2}}} {{\mu^2}\over
{2T \omz}}{\rm cos}(2q t+\beta)
\end{equation}
where $f_0^\prime \equiv \partial{f_0}/\partial{x}$ and tan$\beta
=-(2q)/\gamma$.
The occupation numbers thus `lag' by a phase $\beta$
behind the equilibrium value as that oscillates in time.  As
$\gamma \rightarrow \infty$, this lag goes to zero and $f_{\vec{k}}(t)
\rightarrow f_0\left({{E(\vec{k},t)}\over T}\right)$; the system remains
in equilibrium.  As $\gamma \rightarrow 0$, the interactions turn off
and there is no time variation of $f_{\vec{k}}(t)$.  Maximum dissipation
occurs for $\gamma = 2q$.  The average power dissipated
over a period $(2q)^{-1}$by the mode with momentum $\vec{k}$ is given by
\begin{eqnarray}
P &=& -(2q) \int_0^{(2q)^{-1}} dt \dot{f}_{\vec{k}}(t)
E(\vec{k},t) \nonumber \\
&=& f_0^\prime\left({{\omz}\over T}\right) {{\gamma
(2q)^2}\over {\sqrt{\gamma^2+(2q)^2}}} {{\mu^4}\over
{4 \omz^2 T}} \times \nonumber \\
&~& \int_0^{(2q)^{-1}} dt~ {\rm sin}(2q
t +\beta) {\rm cos}(2q t)
\end{eqnarray}
or
\begin{equation}
P = -f_0^\prime\left({{\omz}\over T}\right)
{{\gamma (2q)^2}\over {\gamma^2+(2q)^2}}
{{\mu^4}\over {8 \omz^2 T}}~.
\end{equation}
The total power dissipated per unit volume is then given by
\begin{equation}
P = -{1 \over 3} \int {{d^3\vec{k}}\over {(2\pi)^3}} f_0^\prime\left(
{{\omz}\over T}\right)
\left[{{\gamma (2q)^2}\over {\gamma^2+(2q)^2}}\right]
{{\mu^4}\over {8 \omz^2 T}}~.
\end{equation}
The factor $1/3$ comes from averaging over all possible
polarizations of the $\chi$ particles.
%(linearly polarized particles, i.e., $A$ or $B$, dissipate as
%given in Eq. (15).  On the other hand, the energy of circularly
%polarized modes does not vary with frequency $2 q$ and thus these
%modes contribute negligibly to the dissipation rate.  The average over
%all polarizations gives $1/3$.)

Equating the dissipation power with the change in axion energy density,
$P = \dot{\rho_a}$, and using $\VEV{q^2} = \VEV{(\epsilon
\rho^{1/2}_a/f_b {\rm cos}(m_at))^2} = {1 \over 2} \epsilon^2
{1 \over {f_b^2}} \rho_a$, the energy
in coherent axion oscillations evolves as
\begin{equation}
\dot{\rho_a} \simeq {1 \over {4\pi^2}}
C(m,T) {{2 \epsilon^2 \rho_a \gamma \mu^4} \over
  {\gamma^2 f_b^2 + 2 \epsilon^2 \rho_a}} ~,
\end{equation}
where $C(m,T) =
\int dx {{x^2}\over{c^2+x^2}} f_0^\prime(\sqrt{c^2+x^2})$,
with $c \equiv m/T$ and $x \equiv k/T$.
Let us now make the approximation that whatever dissipation occurs takes
place in less than an expansion period - so that the functions $\gamma$
and $C(m,T)$ may be approximated as constant. Then, taking the integral
$\int dt \approx 0.1 M_{pl}/T^2$, the final axion energy density
is
\begin{equation}
\rho_{a,f} e^{{2 \epsilon^2  \rho_{a,f}} \over {f_b^2 \gamma^2}} =
\rho_{a,i} e^{{2 \epsilon^2 \rho_{a,i}} \over {f_b^2 \gamma^2}}
e^{-{10^{-3} \epsilon^2 \mu^4 M_{pl}} \over {f_b^2 \gamma T^2}}
\end{equation}
This solution embodies a number of special cases, but we are mostly
interested in the cases where we can treat
$(2 \epsilon^2 \rho_a)/ (f_b^2 \gamma^2)$ as small. Then, the
axion density decreases exponentially during one expansion time. Given
that the maximum initial value of $\rho_a$ is approximately
$\Lambda_{QCD}^4$, as long as
$(10^{-3} \epsilon^2 \mu^4 M_{pl})/ (f_b^2 \gamma T^2)>20$, the axion
density will be reduced to acceptable levels.

We take the temperature to be about $\Lambda_{QCD}$.  Dissipation is
fastest if $f_b$ is small.  On the other hand, if $f_b < T$, the
$U(1)_b$ symmetry will be unbroken and the axion will have no goldstone
boson, $b$, to mix with, and this dissipation mechanism will not operate.
So the greatest dissipation should exist for $f_b \sim T \sim
\Lambda_{QCD}$, and
\begin{equation}
\epsilon^2 \left({{\mu}\over {\Lambda_{QCD}}}\right)^4 \left({{\gamma}\over
{\Lambda_{QCD}}}\right)^{-1} \ge 10^{-16}~.
\end{equation}
On the other hand, the dissipation cuts off for $q \ll \mu^2/m$.
So the energy in the axion field cannot be reduced below
\begin{equation}
\rho_{f,min} \approx {{\mu^4 f_b^2} \over {\epsilon^2 m^2}}
\end{equation}
by this
mechanism. Demanding that this be less than $\sim 10^{-8} \Lambda_{QCD}^4$
in order to comfortably solve the axion energy problem gives
$(\mu/\Lambda_{QCD})^4 \le \epsilon^2 10^{-8}$, and from
Eq. (19), $\epsilon^4 (\gamma/\Lambda_{QCD})^{-1} \ge 10^{-8}$.  These
two constraints can easily be satisfied, for example by
$\epsilon \sim 10^{-3}$, $\gamma < 10^{-4} \Lambda_{QCD}$, and
$\mu^2 \approx 10^{-7} \Lambda_{QCD}^2$.

\section{Discussion}

We have established that it is possible to dissipate the primordial energy
density in axions through their interactions with other particles. A key
feature of the process we envision is to mix the axion with some other
goldstone boson $b$, whose decay constant $f_b$ is much smaller than $f_a$.
The mixing may be large, and this allow axions to couple strongly to `$b$
matter'. By adjusting the matter content of the $b$-sector we can arrange
for the axion oscillations to drive oscillations in the energy of
$b$ matter. This pushes the $b$ matter out of thermal equilibrium, and
ultimately allows the axion energy density to dissipate in the $b$-sector.

Although, we have constructed a model whereby the axions dissipate, the
addition of new particles may have other troublesome consequences, chief
of which is that $b$ matter may itself come to dominate the universe.
The easiest way to avoid this is to adjust the masses and couplings of the
$b$-sector so the real part of the $\Phi$ field
is the lightest $b$-particle other than the $b$ goldstone boson itself.
Then as the universe cools, first all $b$ matter will end up as $\rho_b$'s
which will then decay with a rate of order $f_b$ into $b$ goldstone
bosons.

We argued earlier that it is difficult to dissipate axion energy through
the process of single axion absorption, yet, in the present scenario
axion energy is indeed absorbed. It is natural to ask if there is a
``Feynman diagram'' explanation of our process. The answer is yes.
Imagine a process whereby a single axion is absorbed from the coherent
state, as in Fig. (3a) and
a second process whereby two axions are absorbed, as in Fig. (3b). The
ratio of the rates for these two processes is
\begin{equation}
R = \Gamma_2/\Gamma_1 \sim n_a
  {{\epsilon^2 m_a^2} \over {f_b^2}} ~,
\end{equation}
where $n_a$ is the occupation number for the state.
%and $\Lambda$ is a
%typical energy scale for the medium but depends in detail on the
%interactions of the particle propagating between the two axion vertices.
%The factor of $m_a^2$ assumes derivative couplings.
As stated earlier the
effective single axion absorption rate is reduced by a factor of $m_a /T$
when back reactions are included. However, this is also true of the two
axion absorption rate, and so the ratio in Eq. (21) is correct even in the
presence of backreactions. Now, for the coherent state, $n_a$ is
approximated by $n_a \sim f_a^2/H^2$, so the ratio becomes
$R \sim \epsilon^2 M_{pl}^2/\Lambda_{QCD}^2$, which is quite large.
This leads to the situtation that two axion absorption formally exceeds one
axion absorption. The three axion rate would be even larger... In this
situation what one must do is sum over diagrams with any number of
axions attached to the particles participating in the scattering
process. This is equivalent to solving for the propagator of these
particles in the presence of the coherent oscillating axion field, as we
have done in this paper.

Along these lines, even in a normal axion model with no
mixing each additional axion brings a factor of $M_{pl}/f_a$ to the
amplitude, so here also one should not calculate individual absorption
rates. Rather, one should calculate the propagators for the other
particles, including the time dependent axion field, to arrive at the axion
dissipation rate. We are presently considering the question of
what dissipation may result from the non-derivative
couplings to mesons, and plan to present results in
a separate paper. Although derivative couplings may seem important too,
one must remember that such couplings are less effective at damping the
non-relativistic coherent axion oscillations due to the factor
of $m_a$ in the coupling.

Besides damping the coherent oscillations in the early universe, the
kind of mixing of axions with other goldstone bosons we have been
discussing could lead to other interesting effects.  In particular, it
is possible now to contemplate that even invisible axions with
$f_a \sim M_{GUT}$ or $M_{Pl}$ can have sizable couplings to some kinds
of matter, perhaps even to particles that carry standard model gauge
charges.  This would mean that the ``invisible axion'' may not be quite
as invisible as was thought.

Also worth further investigation is whether similar mixing in familiar
majoron, familon or texture models could lead to interesting phenomena.

\section*{Appendix}

\appeqs

The easiest way to find the eigenstates of the Lagrangian in Eq.~(11) is
to make the field redefinition $\psi = e^{i q t} \chi$, after which the
Lagrangian becomes
\begin{equation}
{\cal L}' = {1 \over 2} \left( \abs{\partial_\mu \psi}^2 - (m^2 -
q^2)\abs{\psi}^2
+ iq(\dot{\psi} \psi^{*} - \psi \dot{\psi}^{*} -
 {1\over2}\mu^2(\psi^2 + \psi^{*2}) \right) ~.
\end{equation}
The equations of motion for ${\cal L}'$ are
\begin{eqnarray}
\ddot{\psi} - \grad^2\psi + (m^2 - q^2)\psi - 2iq\dot{\psi} + \mu^2 \psi^{*} &
= &
  0 \nonumber \\
\ddot{\psi}^{*} - \grad^2\psi^{*} + (m^2 - q^2)\psi^{*} + 2iq\dot{\psi}^{*} +
    \mu^2 \psi & = &  0 ~.
\end{eqnarray}
The solutions to these equations are
\begin{eqnarray}
\psi_i & = & \alpha_i \left(
  e^{i(\omega_+ t - k\cdot x)} + \delta_+ e^{-i(\omega_+ t - k\cdot x)}
\right)
\nonumber \\ & {\rm or} & \nonumber \\
& = & \alpha_i \left(
  e^{-i(\omega_- t - k\cdot x)} + \delta_- e^{i(\omega_- t - k\cdot
x)}\right)
{}~,
\end{eqnarray}
where the index $i$ denotes one of four solutions for a given wave vector
$\vec{k}$. Due to the mixing of $\psi$ and $\psi^{*}$,
to simultaneously solve both equations of motion requires that $\alpha_i$
be either pure real or pure imaginary, but otherwise the normalization
amplitude is arbitrary. The positive eigenfrequencies are given by
\begin{equation}
\omega_\pm  \equiv \left[\omz^2 + q^2
  \pm \left[4 \omz^2 q^2 +\mu^4\right]^{1/2}\right]^{1/2} ~.
\end{equation}
Negative frequency solutions for momentum $\vec{k}$ are equivalent to
positive frequency solutions with momentum $-\vec{k}$, and so are not
distinct. For the case when $\alpha_i$ is real, the mixings are given by
\begin{equation}
\delta_\pm = {{-\mu^2} \over {\omz^2 - (\omega_\pm \pm q)^2}}~.
\end{equation}
For the case where $\alpha_i$ is imaginary, the sign of $\delta$ is changed.

Counting $\omega_\pm$, and the real and imaginary $\alpha_i$, there are
eight eigenfunctions combining the wavenumbers $k$ and $-k$, which is
the right number to determine the field and its first time derivative at
each point; so this is a complete set of states for the $\psi$ field.

For the $\chi$ field, the states are given by $\chi_i = e^{-iqt}\psi_i$.
We can choose any four of the eight eigenstates to characterize the
momentum $k$ (using the other four for $-k$). It seems natural to chose a
set where $\omega - q \rightarrow \omz$ as $\mu^2 \rightarrow 0$, since
in that limit we must get back the free field theory for $\chi$.
With this in mind, the general solution that we will associate with
momentum $k$ is
\begin{eqnarray}
\chi = \sum_{s=\pm} \left(\alpha_{s+}\chi_{s+} + i \alpha_{s-}\chi_{s-}
\right)
\end{eqnarray}
where the $\alpha_{ss^\prime}$ are real numbers and where
\begin{eqnarray}
\chi_{s s^\prime} \equiv N_{ss^\prime}^{-1/2}\left[\left(1-\delta_{ss^\prime}
\right)e^{i(\omega_s-q)t-i k\cdot x} +
\delta_{ss^\prime} e^{-i(\omega_s+q)t+i k\cdot x}\right]~.
\end{eqnarray}
Here
\begin{eqnarray}
N_{ss^\prime} &\equiv & (1-\delta_{ss^\prime})^2 + (\delta_{ss^\prime})^2
\nonumber \\
\omega_s & \equiv & s\left[\omz^2 + q^2 +s\left[4\omz^2
q^2 +\mu^4\right]^{1/2}\right]^{1/2} \nonumber \\
\delta_{ss^\prime} & \equiv & {{\omz^2+s^\prime \mu^2 -(q-
\omega_s)^2}\over {4q \omega_s}}~.
\end{eqnarray}
For $\mu^2 < mq$,
which is the limit taken in the text, Eqs. (A8) reduce to
\begin{eqnarray}
\omega_s \simeq sm+q
\end{eqnarray}
and
\begin{eqnarray}
\delta_{ss^\prime} \simeq {{ss^\prime \mu^2}\over {4 q (m+sq)}
} - {{\mu^4}\over {16 q^2 (m+sq)^2}}~.
\end{eqnarray}
The energy density computed by just substituting Eq. (A6) into the
Hamiltonian density is
\begin{eqnarray}
\rho = \rho_0 + Re\left(\rho_{2q} e^{2iq t}\right) + ...
\end{eqnarray}
where the ellipses represent more rapidly varying terms (eg. terms
varying with frequency $\omega_+-\omega_-$ or $2\omega_s$) and
\begin{eqnarray}
\rho_0 = \left[\sum_{s,s^\prime}(\alpha_{s s^\prime})^2\right]
\omz^2 ~,\nonumber \\
\rho_{2q} = \mu^2 \left(\alpha_{++} + i \alpha_{+-}\right)\left(
\alpha_{-+}+i\alpha_{--}\right)~.
\end{eqnarray}
The approximately `linearly polarized' modes with arg$(\chi) \simeq
\theta/2$ correspond to $(\alpha_{-+}+i\alpha_{--})/(\alpha_{++}+i
\alpha_{+-}) = e^{i\theta}$, whereas the approximately `circularly
polarized' modes correspond to the cases $(\alpha_{++}+i\alpha_{+-})=0$
and $(\alpha_{-+}+i\alpha_{--})=0$.

For linearly polarized modes
\begin{eqnarray}
\rho \propto \left(\omz^2+{1\over 2}\mu^2{\rm cos}(2q t)\right)
\end{eqnarray}
and the energy per quantum is
\begin{eqnarray}
E = \omz + {1 \over 2} {{\mu^2}\over {\omz}}{\rm cos}(2q t)
\end{eqnarray}
as given in the text.

\section*{References}
\begin{enumerate}

\item {R.D. Peccei and H. Quinn, Phys. Rev. Lett. {\bf 38}, 1440 (1977);
\newline S. Weinberg, Phys. Rev. Lett. {\bf 40}, 223 (1978);
\newline F. Wilczek, Phys. Rev. Lett. {\bf 40}, 271 (1978).}
\item {Y. Chikashige, R.N. Mohapatra and R. Peccei, Phys. Lett.
{\bf 98B}, 265 (1981);
\newline G.~Gelmini and M.~Roncadelli, Phys. Lett. {\bf 99B}, 411 (1981);
\newline H.~Georgi, S.~Glashow, and S.~Nussinov, Nucl. Phys. {\bf B193}, 297
(1981).}
\item {F. Wilczek, Phys. Rev. Lett. {\bf 49}, 1549 (1982);
\newline D.B.~Reiss, Phys. Lett. {\bf 115B}, 217 (1982);
\newline G.~Gelmini, S.~Nussinov and T.~Yanagida, Nucl. Phys. {\bf B219},
31 (1983).}
\item {N. Turok, Phys. Rev. Lett. {\bf 63}, 2625 (1989);
\newline N. Turok and D. Spergel, Phys. Rev. Lett. {\bf 64}, 2736 (1990).}
\item {J.E. Kim, Phys. Rev. Lett. {\bf 43}, 103 (1979);
\newline M.A. Shifman, A.I. Vainshtein and V.I. Zakharov, Nucl. Phys.
{\bf B166}, 199 (1981);
\newline M. Dine, W. Fischler and M. Srednicki, Phys. Lett. {\bf 104B}, 199
(1981);
\newline A.P. Zhitniskii, Sov. J. Nucl Phys. {\bf 31}, 260 (1980).}
\item {For reviews, see, J.E. Kim, Phys. Rept. {\bf 149}, 1 (1987);
\newline M.S. Turner, Phys. Rept. {\bf 197}, 67 (1990);
\newline G.G. Raffelt, Phys. Rept. {\bf 198}, 1 (1990).}
\item {J. Preskill, M. Wise and F. Wilczek, Phys. Lett. {\bf 120B}, 127
(1983);
\newline L.F. Abbott and P. Sikivie, Phys. Lett. {\bf 120B}, 133 (1983);
\newline M. Dine and W. Fischler, Phys. Lett. {\bf 120B}, 137 (1983).}
\item{B. Holdom, Phys. Lett. {\bf 166B}, 196 (1986).}
\item{R.L.~Davis, Phys. Rev. {\bf D32}, 3172 (1985);
\newline R.L.~Davis and E.P.S.~Shellard, Nucl. Phys. {\bf B324}, 167
(1989).}
\item{D.~Harari and P.~Sikivie, Phys. Lett. {\bf 195B}, 361 (1987).}
\item{P.~Steinhardt and M.S.~Turner, Phys. Lett. {\bf 129B}, 51 (1983);
\newline G.~Lazarides, \etal, Nucl. Phys. {\bf B346}, 193 (1990).}
\item{S.-Y.~Pi, Phys. Rev. {\bf D24}, 1725 (1984);
\newline A.~Linde, Phys. Lett. {\bf 201B}, 437 (1988).}
\item{J.~Flynn and L.~Randall, LBL preprint, LBL-25115 (1988).}
\item{H.~Georgi, D.~Kaplan, and L.~Randall, Phys. Lett. {\bf 169B}, 73
(1986).}
\end{enumerate}

\newpage
\section*{Figure Captions}
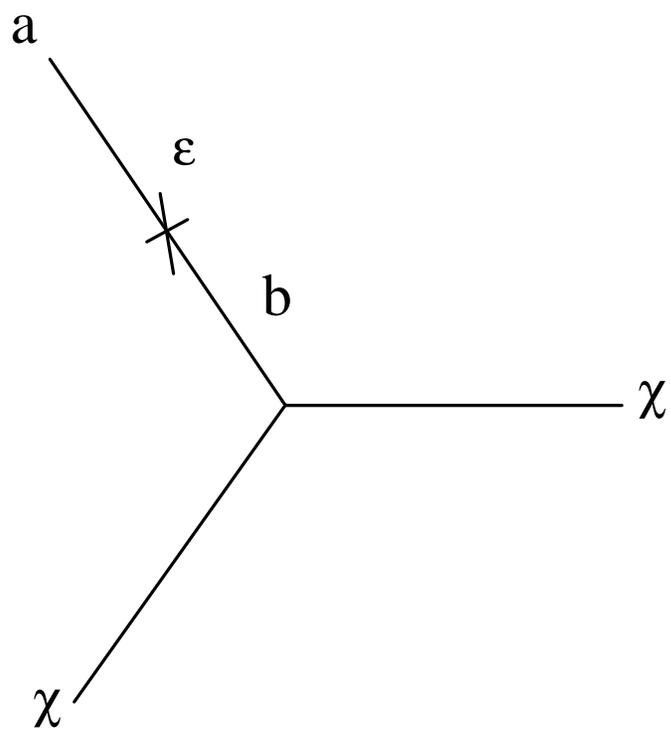
\noindent Figure 1. Effective coupling to `$b$-matter' through goldstone
boson mixing.

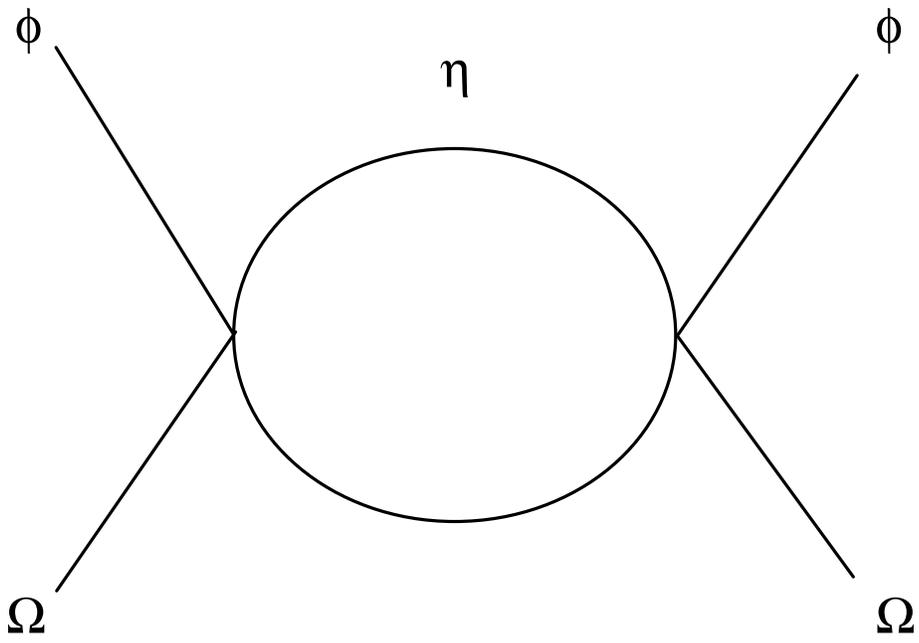
\noindent Figure 2. Mixing of the $\phi$ and $\Omega$ scalars through a
loop. The $\eta$ particle carries both $a$ and $b$ charges.

\noindent Figure 3. Similar processes involving the absorption of
a) a single axion and b) two axions. If the axions come from the coherent
state describing the classical field oscillation, then the rate for
b) exceeds that for a).

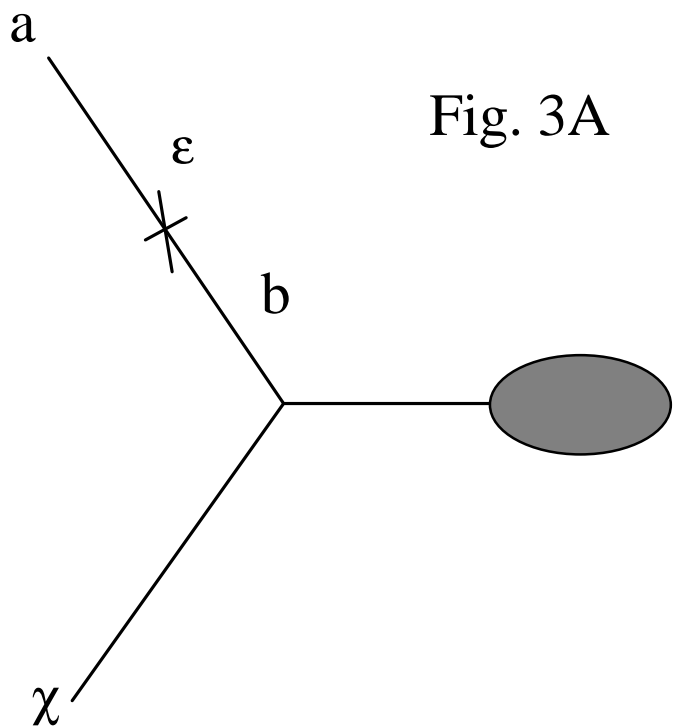
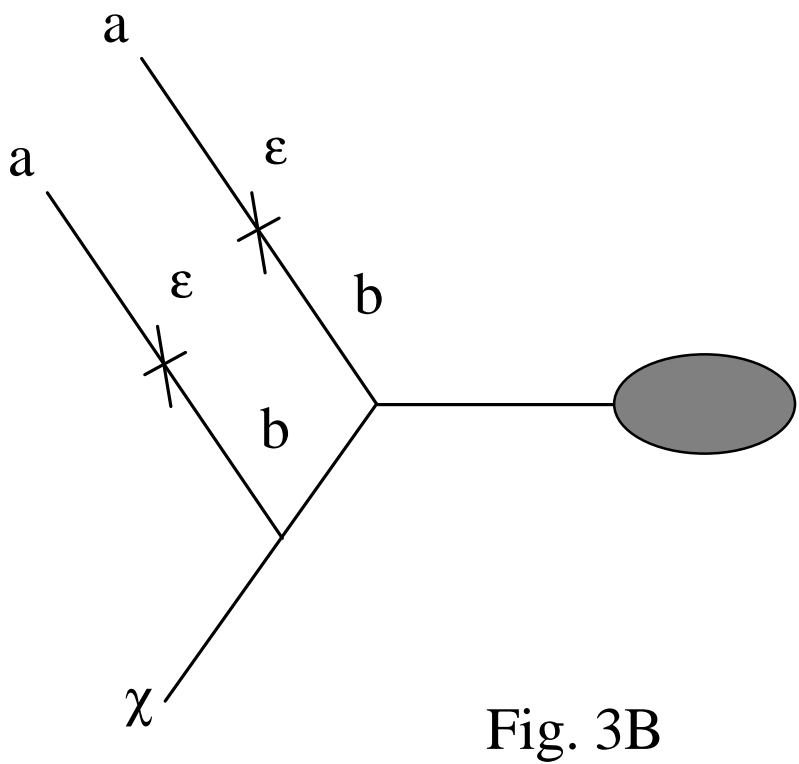
\end{document}